\documentclass[12pt] {article}  

\def\om{\Omega}
\def\wdw{Wheeler--DeWitt}

\def\sch{Schr\"{o}dinger}

\def\be{\begin{equation}}
\def\ee{\end{equation}}

\def\tp1{\tilde p_1}
\def\tp2{\tilde p_2}
\def\tq1{\tilde q^1}
\def\tq2{\tilde q^2}
\def\om{\Omega}

\begin{document}           
\baselineskip.3in

\centerline{\Large\bf Notes on Dilaton Quantum Cosmology\footnote{Chapter 3 of the book {\it Trends in General Relativity and Quantum Cosmology}, Nova Science Publishers, N. Y. (2005).}} 

\medskip

\centerline{\bf Gabriel Catren\footnote{E-mail: catren@iafe.uba.ar}}
\centerline{\it Instituto de Astronom\'{\i }a y
F\'{\i }sica del Espacio}
\centerline{\it C.C. 67, Sucursal 28, 1428 Buenos
Aires, Argentina}
\centerline{and} 
\centerline{\bf Claudio Simeone\footnote{ E-mail:
csimeone@df.uba.ar}}
 \centerline{\it Departamento
de F\'{\i }sica}
\centerline{\it Facultad de Ciencias Exactas y Naturales,
Universidad de Buenos Aires}
\centerline{\it Pabell\'{o}n I, Ciudad Universitaria,
 1428, Buenos Aires, Argentina }        

\vskip1cm

\noindent ABSTRACT: In these notes we address the canonical quantization of the
cosmological models which appear as solutions of the low energy
effective action of closed bosonic string theory (dilaton models).
The analysis is restricted to the quantization of the
mini\-superspace models given by homogeneous and isotropic
cosmological solutions. We  study the different conceptual
and technical problems arising in the Hamiltonian formulation of
these models as a consequence of the presence of the so called
Hamiltonian constraint. In particular we  address the problem
of time in quantum cosmology, the characterization of the symmetry
under clock reversals arising from the existence of a Hamiltonian
constraint, and the problem of imposing boundary conditions on the
space of solutions of the \wdw\ equation.

\newpage

\section{Introduction}

String cosmology received considerable
attention in  the last decade because of the new scenario that it proposes for the early universe  \cite{ven91,ven00,ga00}. When
the high energy modes of the strings become negligible, the
dynamical evolution of the universe begins to be dominated by the
massless fields which act as the matter source of gravitational
dynamics. This phase of the universe is commonly called the {\it
dilatonic era}. The purpose of the present notes is to provide a
consistent quantum description for this epoch. We shall address
the formal aspects of the problem, more precisely, the obtention
of  a wave function allowing for a clear definition of probability
within the context of the minisuperspace approximation.

In the minisuperspace picture all except a few degrees of freedom
of the gravitational and matter fields are frozen at the classical
level, so that the problem to be solved  reduces from quantum
field theory to  quantum mechanics. Most developments in quantum
cosmology have been achieved within the minisuperspace
approximation. However, though the reduction to a finite number of
degrees of freedom considerably simplifies the problem of
obtaining a consistent quantum cosmology, the fact that the
dynamical classical theory includes the general covariance as a
central feature is an obstruction to a straightforward application
of standard quantum mechanics.

This feature is most apparent when the classical theory of the
gravitational field (even with the inclusion of matter) is
formulated in the Hamiltonian form, as one immediately obtains
that the dynamical evolution is governed by a Hamiltonian ${\cal
H}$ which vanishes on the trajectories of the system
\cite{hal90,ba93,si02b}:
$${\cal H}=G^{ik}p_ip_k+V(q)=0.$$
Thus the theory involves a constraint which is quadratic in the
momenta, which reflects the reparametrization invariance of the
corresponding action (see below), i.e., the fact that the
separation between successive spatial hypesurfaces in spacetime is
arbitrary. In general, one also obtains linear constraints,
analogous to those of Yang--Mills theories \cite{ba93}. These
linear constraints assure the invariance of the theory under a
change of the spatial coordinates used to represent the spatial
geometry of each hypersurface. We shall not discuss this point
here, as their role is not central in the minisuperspace picture.

The quantization of constrained systems can be analyzed within
both the canonical approach and the path integral formulation
(see, for example, Ref. \cite{hete92}). In the first of these
frameworks, the direct application to cosmological models of the
well known Dirac program \cite{di64} leads to the \wdw\ equation
\cite{dew}, which is a second order equation in all the
derivatives of the wave function $\Psi$. The solutions to this
equation do not depend explicitly on the ``time'' parameter
$\tau$, but only on the coordinates, which reflects the
reparametrization invariance of the classical theory. This
reparametrization invariance means that the integration parameter
$\tau$ is not a ``true" time. This absence of a notion of time in
the \wdw\ quantization program is a serious obstacle for
understanding the results in terms of conserved positive-definite
probabilities \cite{ha86,befe95,fe99}. Within dilaton cosmology
this problem has received considerable attention (see, for
example, Refs. \cite{cade97,cav99,caun99,caun00,gisi01a}).

The canonical program admits an alternative approach to the Dirac method, which relies on
the idea that it is necessary to ``reduce" the theory before
quantization, i.e., to find a ``true" time at the classical level.
If this reduction can be effectively performed, the original
action can be reduced to an ordinary action without the
reparametrization symmetry of the former. In that case the
quantization program continues as if we were dealing with an
ordinary classical theory. The theory can be then quantized by
means of a \sch\ equation with its typical conserved
positive-definite probabilities.

Our aim will be then the analysis of these points by studying the
cosmological models which appear as solutions of the low energy
effective action of closed bosonic string theory \cite{gave,ga00}.
In Section 2 we shall begin by reviewing dilaton gravity, in
particular homogeneous and isotropic cosmological solutions, whose
formulation we shall put in Hamiltonian form. Then in Section 3 we
shall address the problem of time and quantization, both in the
usual \wdw\ scheme as well as in the \sch\ formulation, and we
shall carefully discuss the relation between the corresponding
solutions, their possible equivalence, and their role in selecting
solutions with physical meaning (in particular when an extrinsic
time must be introduced). Section 4 will be devoted to a more
conceptual than technical discussion, with a wider scope which
goes beyond the particular problem of dilaton models. Finally, in
Section 5 we briefly summarize the essential points of the whole
discussion.

\section{Bosonic string theory  and cosmological models}

\subsection{General theory}

The action describing the world-sheet dynamics of strings on a
curved manifold in presence of background fields has the form \cite{polya,polch}
\begin{eqnarray}
S_{WS} & = &  {1\over 4\pi \alpha '}\int d\sigma d\tau \sqrt{h} \left(
h_{\alpha \beta } g_{\mu \nu }(X)+i\varepsilon _{\alpha \beta }B_{\mu \nu
}(X)\right) \partial ^\alpha X^\mu \partial ^\beta X^\nu \nonumber\\
& & +{1\over 2\pi }\int d\sigma d\tau \sqrt{h}R(X)\phi (X),\label{51}
\end{eqnarray}
where $h_{\alpha \beta }$ is the metric on the string world-sheet,
$R$ is the Ricci scalar related with this metric, $g_{\mu \nu }$
is the metric of the spacetime on which the theory is formulated,
$B_{\mu \nu }$ is an antisymmetric field (commonly known as the  ${NS}$-${NS}$ two-form field) and
$\phi $ is the scalar dilaton field. These three fields appear in
the massless spectrum of closed bosonic string theory. We have noted
 $\alpha ,\beta$ for  the indices corresponding to the coordinates on
the two-dimensional world-sheet, while the indices $\mu ,\nu
,\rho$ correspond to the $D$-dimensional spacetime coordinates.
The parameter $\alpha'$ (commonly known as the Regge slope) is the
inverse of the string tension, $T= 1/(2\pi \alpha')$, which
defines the scale of the theory at the quantum level. Clearly, the
action (\ref{51}) defines a two-dimensional field theory; this
theory is invariant under the transformations \be \delta B_{\mu
\nu } =\partial _\mu \Lambda _\nu -\partial _\nu \Lambda _\mu, \ \
\ \ \ \ \ \ \delta \phi =\phi _0,  \label{52} \ee with $\Lambda
_\mu $ an arbitrary vector and $\phi _0$ a constant.

The action (\ref{51}) is invariant under the  Weyl --conformal--
transformation $h_{\alpha \beta }\to\Omega^2(\tau,\sigma)h_{\alpha
\beta }$. If we require that this symmetry holds also at the
quantum level (no conformal anomalies) the beta functions
 must vanish \cite{polch,ga00}; then  at
first order in the $\alpha'$ power expansion and introducing the
strength tensor $H_{\mu \nu \rho }$ associated to the
antisymmetric field $B_{\mu \nu }$
\begin{equation}
{\bf  H}_{\mu \nu \rho }=\partial _\mu B_{\nu \rho }+\partial
_\rho B_{\mu \nu }+\partial _\nu B_{\rho \mu }, \label{53}
\end{equation}
 we obtain:
\begin{eqnarray}
R_{\mu \nu }+\nabla _\mu \nabla _\nu \phi -\frac 14{\bf  H}_{\mu \rho
\delta }{\bf  H}_\nu ^{  \rho \delta } & = & 0, \nonumber\\
\nabla ^\delta {\bf  H}_{\delta \mu \nu }-\nabla ^\delta \phi {\bf  H}%
_{\delta \mu \nu } & = & 0, \nonumber\\
c-\nabla _\mu \nabla ^\mu \phi +\nabla _\mu \phi \nabla ^\mu \phi -\frac 16%
{\bf  H}_{\mu \nu \rho }{\bf  H}^{\mu \nu \rho } & = & 0,
\label{54}
\end{eqnarray}
where, in principle, $c= 2(D-26)/(3\alpha')$. However, $c$ can be
changed by including more fields, so that in what follows we shall
consider it as an arbitrary real number. Note then the difference
between the theory for a point particle and the theory for a
string: the quantum theory for the last one can not be
consistently formulated in an arbitrary background, but, instead,
it imposes restrictions on the admissible external fields.

Now, we are interested in a spacetime formulation of the theory of
gravitation, analogous to the Einstein--Hilbert action that we
have in General Relativity.  It can be shown that the equations (\ref{54}) can be interpreted as the Euler--Lagrange
equations of motion of a field theory corresponding to the
following effective action:
\begin{equation}
S_{SF}=\frac 1{16\pi G_N}\int d^Dx\sqrt{-g}e^{-\phi }\left( -c+R+\nabla _\mu
\phi \nabla ^\mu \phi -\frac 1{12}{\bf H}_{\mu \nu \rho }{\bf H}^{\mu
\nu \rho }\right),  \label{be}
\end{equation}
where $G_N$ is the $D$-dimensional Newton constant and $R$ is the
Ricci scalar of the spacetime. Thus, we can understand this as the
low energy effective action describing the large tension limit
($\alpha'\rightarrow 0$) of closed bosonic string theory. A
consistent configuration of background fields for the formulation
of string theory must be a classical solution obtained from the
variational principle corresponding to the action (\ref{be}). In
brief, the requirement of preserving at the quantum level the
conformal invariance of the two-dimensional world sheet theory,
formulated up to the first order in the inverse of the string
tension, leads to the same equations of motion resulting from the
variational principle $\delta S=0$ imposed on the $D$-dimensional
field theory given by (\ref{be}).

A more familiar formulation can be obtained by redefining the
fields as $ g_{\mu \nu }\rightarrow e^\phi g_{\mu \nu }, $ so that
the action of the spacetime theory becomes
\begin{eqnarray}
S_{EF} & = & {1\over 16\pi G_N}\int d^Dx\sqrt{-g}\nonumber\\
& & \times\left( R-ce^ {2\phi/(D-2)}
+{1\over D-2}\nabla _\mu \phi \nabla ^\mu \phi -{e^{-4\phi/(D-2)}\over 12}{\bf
H}_{\mu \nu \rho }{\bf H}^{\mu \nu \rho }\right).
\end{eqnarray}
Thus we have obtained  the $D$-dimensional Einstein action
including coupling terms with  the dilaton and the antisymmetric
field. This form for the effective field theory is known as {\it
Einstein frame action}, while  (\ref{be}) is commonly called the
{\it string frame action}. The interpretation of the gravitational
aspects of the theory is more clear in the Einstein frame: in the
particular case $D=4$, the variational principle $\delta S=0$
imposed to this new form of the effective action leads to the
equations
\begin{equation}
\nabla_\mu\partial^\mu \phi +ce^\phi -{1\over 16}e^{-2\phi }{\bf
H}^2=0,  \label{58}
\end{equation}
\begin{equation}
\nabla _\delta {\bf H}_{  \mu \nu }^\delta +2\nabla _\delta \phi
{\bf H}_{  \mu \nu }^\delta =0,  \label{59}
\end{equation}
\begin{eqnarray}
R_{\mu \nu }-{1\over 2} g_{\mu \nu }R-{c\over 2}g_{\mu \nu }e^\phi & = &
{1\over 2}\left( \nabla _\mu \phi \nabla _\nu \phi -{1\over 2}g_{\mu \nu }(\nabla
\phi
)^2\right) +  \nonumber \\
& &+{1\over 4}e^{-2\phi }\left( {\bf H}_{\mu \rho \delta }{\bf H}_\nu ^{%
  \rho \delta }-{1\over 6}g_{\mu \nu }{\bf H}^2\right),
\label{510}
\end{eqnarray}
and the Bianchi identities
\begin{equation}
\nabla _{[\mu }{\bf  H}_{\mu \rho \delta ]}=0.  \label{511}
\end{equation}
We can recognize in (\ref{510}) the Einstein equations with a
cosmological function given by $\Lambda =ce^\phi$ and  with the
energy-momentum tensor of the dilaton and the antisymmetric fields
as the source. The relation between the string frame and the
Einstein frame formulations can be clarified by considering a
solution of the equations (\ref{be}) in the case of homogeneity
and isotropy. Because the metric $ds^2$ in the string frame is
related to the corresponding metric $d\tilde s^2$ in the Einstein
frame by $ds^2=e^\phi d\tilde s^2$, one can easily translate the
results. For example, for the case $c=0$ a possible solution is a
flat cosmology with a metric  $ds^2$ where the scale factor
behaves like $a\sim \tau^{1/3}$. By recalling the corresponding
evolution of the dilaton with $\tau$, this behavior is translated
to the Einstein frame as an evolution of the scale factor
$b\sim\tau^{1/2}$, that is, to the same evolution of a
radiation-dominated universe (see \cite{gope} for a detailed
discussion).

\subsection{Cosmological models and Hamiltonian formulation}

Consider a cosmological model with a finite number of degrees of
freedom identified by the coordinates $q^i$ (geometrical and
matter degrees of freedom). The Lagrangian form for the action of
such a minisuperspace is
\begin{equation}
S[q^i,N]=\int _{\tau_1}^{\tau_2} N \left( {1\over 2N^2} G_{ij}{dq^i\over
d\tau}{dq^j\over d\tau}-V(q)\right)d\tau\label{SL}
\end{equation}
where a spatial integration  must be understood in the integrand.
In (\ref{SL}), $G_{ij}$ is the reduced version of the DeWitt
supermetric (see, for example, \cite{ba93}), $V$ is the potential,
which depends on the curvature and the coupling  between the
fields, and $N(\tau)$ is the lapse function determining the
separation between spacelike hypersurfaces in spacetime
\cite{ba93,hal90,si02b}.

The role of constraints and Lagrange multipliers becomes manifest
in the Hamiltonian formulation, being this formalism the best
suited for the canonical quantization of  cosmological  models. If
we define the canonical momenta as
$$p_i={1\over N}G_{ij}{dq^j\over d\tau},$$
we obtain the Hamiltonian form of the action
\begin{equation}
S[q^i,p_i,N]=\int_{\tau_1}^{\tau_2} \left( p_i{dq^i\over d\tau}-N{\cal
H}\right)d\tau,\label{28}
\end{equation}
where
\begin{equation}
{\cal H}=G^{ij}\,p_ip_j+V(q).
\end{equation}
Under arbitrary changes of the coordinates $q^i$, the momenta
$p_i$ and the lapse function $N$ we obtain
\begin{equation}
\delta S  =  \left. p_i\delta
q^i\right|_{\tau_1}^{\tau_2}+\int_{\tau_1}^{\tau_2}\left[\left({dq^i\over
d\tau}-N{\partial {\cal H}\over\partial p_i}\right)\delta
p_i-\left({dp^i\over d\tau}+N{\partial {\cal H}\over\partial
q_i}\right)\delta q^i-{\cal H}\delta N\right]d\tau.\end{equation}
Then if we demand the action to be stationary when the coordinates
$q^i$ are fixed at the boundaries,  we obtain on the classical
path the Hamilton canonical equations
\begin{equation}
{dq^i\over d\tau}=N[q^i,{\cal H}],\ \ \  \ \ \ \ \    {dp_i\over d\tau}= N[p_i,{\cal
H}]
\end{equation}
and the Hamiltonian constraint
\begin{equation}
{\cal H}=0.
\end{equation}

Two features of the dynamics can thus be remarked. The first one
is that the presence of the constraint ${\cal H}=0$ restricts
possible initial conditions to those lying on the constraint
surface. The second is that the evolution of the lapse function
$N$ is arbitrary, i.e., it is not determined by the canonical
equations; hence, the separation between two successive
three-surfaces is arbitrary, which constitutes  the
minisuperspace version of the many-fingered nature of time of the
full theory of gravitation \cite{ba93,si02b}.

The field equations yielding from the spacetime action (\ref{be})
admit homogeneous and isotropic solutions in four dimensions
\cite{ant88,tsey92a,tsey92b,gope}. Such solutions have a metric of
the Friedmann--Robertson--Walker form:
\begin{equation}
ds^2=N(\tau )d\tau ^2-a^2(\tau)\left( \frac{dr^2}{1-kr^2}+r^2d\theta
^2+r^2\sin ^2\theta d\varphi ^2\right),  \label{frw}
\end{equation}
where  $a$ is
the scale factor and $k=(-1,0,1)$ determines the curvature. For
the dilaton $\phi$ and the field strength $H_{\mu \nu \rho }$ the
homogeneity and isotropy requirements lead to \be \phi
=\phi (\tau) \ \ \ \ \ \ \ \ {\bf H}_{ijk} =\lambda
(\tau)\varepsilon _{ijk}\label{513} \ee where $\varepsilon _{ijk}$
is the volume form on the constant-time surfaces and $\lambda$ is
a real number. The Bianchi identities (\ref{511}) imply that
$\lambda $ does not depend on the parameter $\tau$. An important aspect of these cosmological solutions is that they allow for a conceptually new scenario for the early universe, where the standard big bang is replaced by a phase of finite curvature \cite{gave,ga00}. 

Let us write down the explicit form of the action for the models
to be considered here. If we define $b^2(\tau)\equiv
e^{2\Omega(\tau)}$, the Lagrangian form of the Einstein frame
action in four dimensions for the case $\lambda=0$, which
corresponds to the two-form field $B_{\mu\nu}$ equal to zero, is
given by
\begin{equation} S={1\over 2}\int_{\tau_1}^{\tau_2} d\tau
Ne^{3\Omega}\left[-{{\dot\Omega}^2\over N^2}+{{\dot\phi}^2\over
N^2}-2ce^\phi+ke^{-2\Omega}\right],
\end{equation}
where a dot stands for $d/d\tau$. On the other hand,
in the case $k=0$  (flat universe) we can write
\begin{equation}
 S={1\over 2}\int_{\tau_1}^{\tau_2} d\tau Ne^{3\Omega}\left[-{{\dot\Omega}^2\over
N^2}+{{\dot\phi}^2\over N^2}-2ce^\phi-\lambda^2e^{-6\Omega-2\phi}\right].
\end{equation}
We have absorbed the factor $(8\pi G_N)^{-1}$ by redefining the fields.

The Hamiltonian form of the Einstein frame action for the models considered
reads
\begin{equation}
 S=\int_{\tau_1}^{\tau_2} d\tau \left[p_\Omega\dot\Omega+p_\phi\dot\phi-N{\cal
H}\right].
\end{equation}
 For
$\lambda=0$ the Hamiltonian constraint is
\begin{equation}
{\cal  H}={1\over 2}e^{-3\Omega}\left(-p_\Omega^2+p_\phi^2+2c
e^{6\Omega+\phi}-ke^{4\Omega}\right)= 0,\label{519}
\end{equation}
while for $k=0$ we have
\begin{equation}
{\cal  H}={1\over 2}e^{-3\Omega}\left(-p_\Omega^2+p_\phi^2+2c e^{6\Omega+\phi}+
\lambda^2 e^{-2\phi}\right)= 0.\label{520}
\end{equation}
These two Hamiltonian constraints, or their scaled forms $H\equiv
2e^{3\Omega}{\cal H}=0$ (see Appendix A), will be the starting
point for  the canonical quantization of the models.

\section{Quantization}

\subsection{\sch\ and \wdw\ equations}

The standard procedure for quantizing the minisuperspaces models
described by the Hamiltonian constraints (\ref{519}) and
(\ref{520}) is to turn them into operators and make them act on a
wave function (Dirac method). This prescription yields the usual
Wheeler--DeWitt equation, which is an hyperbolic equation of
second order in all its derivatives. However, the absence of a
true time in the classical formalism (which is reflected in the
appearance of the Hamiltonian constraint ${\cal H}=0$) makes
difficult the interpretation of the resulting wave function in
terms of a conserved positive-definite inner product. One knows in
fact how to define a conserved positive-definite inner product for
the solutions of a Schr\"{o}dinger equation,  and not for the
solutions of a Klein--Gordon type equation as the Wheeler--DeWitt
equation. Other problems associated with the Wheeler--DeWitt
equation are discussed in \cite{SSV1,SSV2,taty}.

However, in certain cases it is possible, by means of the
identification of a time variable among the canonical variables,
to extract from the time independent Wheeler--DeWitt equation a
time dependent Schr\"{o}dinger equation. In this case a clear
probability interpretation can be given to the space of solutions
of the Schr\"{o}dinger equation. As Barbour said \cite{barb99}, in
canonical quantum gravity the situation is in a certain sense
inverted with respect to ordinary quantum mechanics where the time
independent Schr\"{o}dinger equation $\widehat{h} \psi=E \psi$ is
derived from the more general time dependent Schr\"{o}dinger
equation $\widehat{h} \psi=i \frac{\partial \psi}{\partial t}$. In
canonical quantum gravity a time dependent Schr\"{o}dinger
equation has to be extracted from a time independent
Schr\"{o}dinger equation with autovalue zero ${\widehat{\cal H}
\psi}=0$ (Wheeler--DeWitt equation).

There are mainly two possibilities for defining a Schr\"{o}dinger
equation for a dynamical system with a constraint ${\cal H}=0$
quadratic in all its momenta. The first possibility it to perform
a canonical transformation which transforms the quadratic
Hamiltonian constraint in a constraint linear in one of its
canonical momenta. If a certain positivity condition is satisfied
by this linear momentum, its canonically conjugated coordinate can
then be defined as the ``real'' time parameter for the evolution
of the system \cite{befe95,fe99,ku95}. 

The other possibility is to factorize the scaled Hamiltonian
constraint $H\equiv 2e^{3\Omega}{\cal H}=p_{0}^{2}-h^{2}=0$ (with
$h^{2}=p_{\mu}p^{\mu}+V \left(q^{\mu}, q^{0} \right)$) in two
disjoint sheets $H=\left(p_{0}+h \right)\left(p_{0}-h \right)=0$,
given by the two signs of the momentum $p_0$ conjugated to the
coordinate $q^0$ that one wants to identify with time (modulo a
sign). Below we shall analyze the conceptual meaning of this
factorization. A time dependent Schr\"{o}dinger equation can then
be associated to each factor $p_{0} \pm h$. The splitting of the
original constraint in two constraints requires the existence of a
non vanishing momentum $p_{0}$ (a \emph{true} time must not invert
its sense of evolution), which may not happen in the original
variables describing the model. Indeed, for $q^0$ to be a right
time choice, the condition $[q^0,{\cal H}]>0$ must be fulfilled
(see \cite{ha86} and the Appendix A), leading this condition to
the requirement that $p_0\neq 0$. Besides, for obtaining a proper
quantum theory in each sheet, the operator corresponding to the
reduced Hamiltonian $h$ must be a self-adjoint operator in order
to have a unitary evolution. Given that in the factorization
process the reduced Hamiltonian $h$ is obtained by taking a square
root $\left(h=\sqrt{p_{\mu}p^{\mu}+V \left(q^{\mu}, q^{0}
\right)}\right)$, the associated operator $\hat{h}$ will be
self-adjoint only if the term under the square root is positive
defined, being this fact an important constraint to the reduction
process that we have described. It may also happen that the model
can not be reduced in the original canonical variables, being then
necessary to perform a canonical transformation in order to
properly reduce the system with a self-adjoint Hamiltonian
operator $\hat{h}$. In that case it is possible that the selected
time variable depends both on the original canonical coordinates
and momenta ($t \equiv q^{0}= q^{0} \left(q^{\mu}, p_{\mu}
\right)$), being these kind of time variables called
\emph{extrinsic times} \cite{ku71,yo72} (in contrast to the
\emph{intrinsic times} which depend only on the canonical
coordinates $q^{\mu}$).

If the reduction process can be successfully performed, then we
will have a pair of Hilbert spaces, each one with its
corresponding Schr\"{o}dinger equation of first order in
$\partial/\partial q^0$ with a reduced Hamiltonian operator
$\hat{h}$. The physical inner product is defined in each space as
\be (\Psi_2|\Psi_1)=\int dq \,\delta(q^0-\tilde
q^0)\Psi_2^*\Psi_1. \ee

We could say that the Schr\"{o}dinger quantization preserves the
topology of the constraint surface: the splitting of the classical
solutions into two disjoint subsets has its quantum version in the
splitting of the theory in two Hilbert spaces \cite{car94,si03}.

\subsection{Intrinsic time}

We shall begin by considering a generic (scaled) Hamiltonian constraint of the
form
\begin{equation}
- p_1^2+ p_2^2+Ae^{(a q^1+b q^2)}=0, \label{613}
\end{equation}
with $a\neq b$. This constraint admits an intrinsic time because
the potential does not vanish for finite values of the coordinates
(see Appendix A; for models not fulfilling this condition see the
next section). This Hamiltonian constraint corresponds to several
dilaton cosmologies, namely, to the cases $\lambda=0,k=0,c\neq 0$;
$\lambda=0, k=\pm 1, c=0$; $\lambda\neq 0, k=0, c=0$ (see Ref.
\cite{gisi01a} for the quantization of the corresponding
minisuperspaces within the path integral formulation). It is easy
to show that the coordinate change
\begin{equation}
x \equiv  {1\over 2}\left(a\Omega+b\phi\right)\  \ \ \ \ \ \
y \equiv  {1\over 2}\left(b\Omega+a\phi\right)
\end{equation}
leads to the following form of the constraint:
\begin{equation}
H=-p_x^2+p_y^2+\zeta e^{2x}= 0, \label{927}
\end{equation}
with $sgn (\zeta)=sgn (A/(a^2-b^2))$. The Wheeler-DeWitt equation
corresponding to this constraint is
\begin{equation}
\left( \frac{\partial^{2}}{\partial x^{2}} -
\frac{\partial^{2}}{\partial y^{2}} + \zeta e^{2x}\right) \Psi
_\omega (x,y) = 0.\label{735}
\end{equation}
The general solution for the case $\zeta >0$ is
\begin{eqnarray}
\Psi _\omega (x,y) & = &\left[ a_{+}(\omega )e^{i\omega
y}+a_{-}(\omega )e^{-i\omega y}\right]\nonumber\\
& & \times \left[ b_{+}(\omega )J_{i\omega }(\sqrt{\left| \zeta
\right| }e^x)+b_{-}(\omega )N_{i\omega }(\sqrt{\left| \zeta
\right| }e^x)\right]\label{tt},
\end{eqnarray}
with $J_{i\omega} $ and $N_{i\omega}$ the Bessel and Neumann
functions of imaginary order respectively. Instead, for $\zeta
<0$, the general solution is
\begin{eqnarray}
\Psi _\omega (x,y) & = & \left[ a_{+}(\omega )e^{i\omega
y}+a_{-}(\omega )e^{-i\omega y}\right]\nonumber\\
& & \times\left[ b_{+}(\omega )I_{i\omega }(\sqrt{\left| \zeta
\right| }e^x)+b_{-}(\omega )K_{i\omega }(\sqrt{\left| \zeta
\right| }e^x)\right],  \label{ttt}
\end{eqnarray}
where $I_{i\omega}$ and $K_{i\omega}$ are  modified Bessel
functions. The obtention of these solutions to the
Wheeler--DeWitt equation does not finish the quantization process
given that we do not know yet which are the proper boundary
conditions that we should impose on these spaces of solutions.

Instead of quantizing the model by solving the Wheeler--DeWitt
equation, one could also reduce the system by identifying a time
variable and solving the corresponding Schr\"{o}dinger equations.
Depending on the sign of the constant $\zeta$ in the constraint
(\ref{927}), these models admit as global phase time the
coordinates $x$ or $y$. In the case $\zeta>0$ the time is $t=\pm
x$, so that we can define the reduced Hamiltonian as
$h=\sqrt{p_y^2+\zeta e^{2x}}$, being this reduced Hamiltonian time
dependent. If, instead, we
have $\zeta<0$, the time is $t=\pm y$ and the reduced Hamiltonian
is $h= \sqrt{p_x^2-\zeta e^{2x}}$. In the first case, $\zeta>0$
($t=\pm x$), the corresponding Schr\"{o}dinger equations are
\begin{equation}
i{\partial\over\partial x}\Psi(x,y)=\pm
\left(-{\partial^2\over\partial y^2}+\zeta
e^{2x}\right)^{1/2}\Psi(x,y)
\end{equation}
In the second case ($\zeta<0$, $t=\pm y$) the associated
Schr\"{o}dinger equations are
\begin{equation}
i{\partial\over\partial y}\Psi(x,y)=\pm
\left(-{\partial^2\over\partial x^2}-\zeta
e^{2x}\right)^{1/2}\Psi(x,y).
\end{equation}
For both $\zeta>0$ and $\zeta<0$ we have a pair of Hilbert spaces,
each one with its corresponding Schr\"{o}dinger equation, and a
conserved positive-definite inner product allowing for the usual
probability interpretation of the wave function (this is analogous
to the  obtention of two quantum propagators, one for each
disjoint theory, in the context of path integral quantization
\cite{si03,ba93}).

It is important to remark that in the present model the time
variables which permit to reduce the system ($t= \pm x$ and $t=
\pm y$ for the cases $\zeta>0$ and $\zeta<0$ respectively) can be
selected among the original canonical coordinates, being then
unnecessary to perform any kind of canonical transformation. In
cases like this one the Schr\"{o}dinger quantization formalism can
be applied directly from the beginning in a straightforward
manner.

It is also important to insist on the fact  that the cases
$\zeta>0$ and $\zeta<0$ are not symmetric: if in the case
$\zeta<0$ ($t= \pm y$) the reduced Hamiltonian $h=h\left(x, p_{x}
\right)$ is time independent, in the case $\zeta>0$ ($t= \pm x$)
the reduced Hamiltonian $h=h\left(x, p_{y} \right)$ is time
dependent.

For the case $\zeta<0$, the stationary solutions to the
Schr\"{o}dinger equations corresponding to both sheets ($t=\pm y$)
are
\begin{equation}
\Psi_{\pm}^{\omega}(x,y)= e^{\mp i \omega y}\left[ b_{+}(\omega
)J_{i\omega }(\sqrt{\left| \zeta \right| }e^x)+b_{-}(\omega
)N_{i\omega }(\sqrt{\left| \zeta \right| }e^x)\right]
\end{equation}
It is clear from this expression that the space of solutions
(\ref{ttt}) of the \wdw\ equation for $\zeta<0$ is the direct sum
of the spaces of solutions of the \sch\ equations corresponding to
each sheet of the Hamiltonian constraint. For the time dependent
case ($\zeta>0, t=\pm x$) the relation between both spaces of
solutions is not so clear due to the presence of operator ordering
ambiguities (this case will be discussed in Section 3.5).

Another point to be remarked is that as a result of the right
definition of time, in both cases the reduced Hamiltonian $h$ are
real, so that the evolution operator is self-adjoint and the
resulting quantization is unitary. Instead, a wrong choice of
time, like for example $t=\pm x$ in the case $\zeta<0$, leads to a
reduced Hamiltonian $h$  which is not real for all allowed values
of the variables, and we obtain a non unitary theory. Another
remarkable aspect is that a right time is not necessarily
associated to the geometrical degrees of freedom, as one could
naively expect. In the case $\lambda\neq 0, k=0, c=0$ for example,
one obtains that the \emph{geometrical} coordinate $\Omega$ is the
physical clock and that the reduced Hamiltonian involves only the
dilaton and the antisymmetric field (the last one, through the
constant $\lambda$). On the contrary, in the case $\lambda= 0,
k=1, c=0$ the physical clock is the dilaton, while the reduced
Hamiltonian depends only on the geometry (see Section 4.1).

\subsection{Extrinsic time}

We shall now consider the problem of  models not admitting a
global time defined only in terms of the coordinates. A good example  is
given by a nontrivial dilaton cosmology described by the scaled
Hamiltonian constraint
\begin{equation}
H= -p_\Omega^2+p_\phi^2+2ce^{6\om+\phi}+\lambda^2e^{-2\phi}= 0,\label{pop}
\end{equation}
which corresponds to a flat universe with dilaton field $\phi$ and
non vanishing antisymmetric field $B_{\mu\nu}$ coming from the
$NS$-$NS$ sector of effective string theory. This model is not solvable by separating variables, and in the
case $c<0$ the potential can vanish, so that it  does not admit an
intrinsic time. However, because  these cosmologies come from the
low energy string theory, which makes sense in the limit $\phi\to
-\infty$, then the $e^\phi\equiv V(\phi)$ factor in the first term
of the potential verifies $V(\phi)=V'(\phi)\ll 1$, and we can
replace $ce^\phi$ by the constant $\overline c$ fulfilling
$|\overline c|\ll |c|$:
\begin{equation}
H= -p_\Omega^2+p_\phi^2+2\overline c e^{6\om}+\lambda^2e^{-2\phi}= 0.\label{popy}
\end{equation}

Following Ref. \cite{moncrief}, we shall start in a naive way  by
quantizing  by means of a Wheeler--DeWitt equation the model
described by the Hamiltonian constraint (\ref{popy}). However, for
this model we can also perform a canonical transformation which
permits to obtain a Hamiltonian constraint that can be factorized
in a two-sheet constraint of the form $H=\left(p_{0}+h \right)
\left(p_{0}-h \right)=0$ \cite{cafe01}. We will then be able to
quantize the model also by using the Schr\"{o}dinger equations
corresponding to each sheet of the factorized Hamiltonian
constraint. It is important to remark that this Schr\"{o}dinger
quantization is not trivial in the sense that a non trivial
canonical transformation is needed. If this canonical
transformation were not known, we should face the problem of
imposing boundary conditions on the space of solutions of the
Wheeler--DeWitt equation. Given that this model can be quantized
by both methods, it is of great value in order to understand how
to impose boundary conditions to the solutions of the
Wheeler--DeWitt equation for cases in which a global time is not
known.

 The Wheeler--DeWitt equation associated to the constraint (\ref{popy}) is
\begin{equation}
\left({\partial^2\over\partial \Omega^2}- {\partial^2\over\partial
\phi^2}+{2\overline
c}e^{6\om}+\lambda^2e^{-2\phi}\right)\Psi(\om,\phi)=0,\label{rty}
\end{equation}
and its  solutions are
\begin{eqnarray}
\Psi_\omega(\om,\phi) & = &
\left[a(\omega)I_{i\omega}\left({|\lambda|}e^{-\phi}\right)+b(\omega)K_{i\omega}\left({|\lambda|}e^{-\phi}\right)\right]\nonumber\\
& & \times\left[c(\omega)I_{i\omega/2}\left(\sqrt{|2\overline
c|}e^{3\om}\right)+d(\omega)K_{i\omega/2}\left(\sqrt{|2\overline
c|}e^{3\om}\right)\right],\label{rtu}
\end{eqnarray}
where $I_{\nu}$ and $K_{\nu}$ are modified Bessel functions. In
principle one does not know which are the proper boundary
conditions that should be imposed on these solutions. If one was
to proceed in a naive way without any consideration about time,
then both contributions including the functions $I_{\nu}$ should
be discarded, because they diverge in what would be commonly
understood as a region classically forbidden by the behavior of
the exponential terms in the potential. In fact, this has been the
choice in the case of the Taub universe in Ref. \cite{moncrief}.
However, as we shall show, the functions $I_{i\omega}(|\lambda |
e^{-\phi})$ should not be discarded, because in the picture
including a globally right notion of time, the dilaton $\phi$ is
associated to the physical clock.

This can be clearly realized by performing the canonical
transformation introduced for the Taub universe in Ref.
\cite{cafe01} in order to obtain a constraint with only one term
in the potential. This is achieved by introducing the generating
function of the first kind
\begin{equation}
\Phi_1 (\phi,s)= \pm|\lambda|e^{-\phi} \sinh s .\label{leo}
\end{equation}
The new canonical variables are then given by
\begin{eqnarray}
s & = & \pm \mathrm{arcsinh}\left({p_\phi  e^{\phi}\over|\lambda|}\right)\nonumber\\
p_s & = & \pm |\lambda|e^{-\phi} \cosh s.
\label{471}\end{eqnarray}
With this canonical transformation the resulting form for the
Hamiltonian constraint in the limit $V(\phi)=V'(\phi)\ll 1$ is
\begin{equation}
H=-p_\om^2+p_s^2+2\overline c e^{6\om}=0,\label{joi}
\end{equation}
and we can  apply our procedure starting from this constraint. In
the case  $\overline c<0$, the momentum $p_s$ does not vanish and
the time is $t=\pm s$. According to the definition of the new
variable $s$ the time $t=\pm s$ is a function of both $p_\phi$ and
$\phi$, being then an {\it extrinsic time}. Once again, observe
that, differing from what is sometimes considered the ``natural''
choice, the physical clock is not associated to the metric, but to
the matter field (see below). The constraint (\ref{joi}) can be
written as
\begin{equation}
H=\left(p_s-\sqrt{p_\om^2-2\overline c e^{6\om}} \right)
\left(p_s+\sqrt{p_\om^2-2\overline c e^{6\om}} \right)=0\label{lse}
\end{equation}
(see Section 3.5). If this constraint equation is satisfied by
demanding $p_s=h=\sqrt{p_\om^2+|2\overline c| e^{6\om}}$, then
this choice implies (taking into account that $p_{t}=-h$) that one
has selected the variable $-s$ to play the role of the physical
clock ($t=-s$ and $p_{t}=-p_{s}$). If on the contrary one demands
$p_s=-h=-\sqrt{p_\om^2+|2\overline c| e^{6\om}}$, then one is
choosing $s$ as the physical clock ($t=s$ and $p_{t}=p_{s}$). In
both cases the positive definite reduced Hamiltonian $h$ does not
depend on the time variable $t=\pm s$. As we will explain later,
the fact that the reduced Hamiltonian $h$ is time independent
assures the equivalence at the quantum level of the factorized
constraint (\ref{lse}) and the original one given by (\ref{joi}).
The corresponding Schr\"{o}dinger equations associated to the
choices $t=s$ and $t=-s$ are respectively
\begin{eqnarray}
i{\partial\over\partial s}\Psi_{+}(\om,s)= \left(-
{\partial^2\over\partial \om^2}+|2\overline
c|e^{6\om}\right)^{1/2}\Psi_{+}(\om,s). \label{682}
\\-i{\partial\over\partial s}\Psi_{-}(\om,s)= \left(-
{\partial^2\over\partial \om^2}+|2\overline
c|e^{6\om}\right)^{1/2}\Psi_{-}(\om,s). \label{683}
\end{eqnarray}
By $\Psi_{\pm}$ we have noted the solutions corresponding to the
sheets $K_{\pm}=p_{s}\pm h=0$. Note that $\Psi_{+}=\left(\Psi_{-}
\right)^{\ast}$ (this relation between both spaces of solutions
will be discussed later). Given that the reduced Hamiltonian
$h=\sqrt{p_\om^2-2\overline c e^{6\om}}$ does not depend on time,
one can propose a solution of the general form $\Psi_{E}(\om,t)=
\varphi_{E} \left(\Omega \right) e^{-iEt}$, where $\varphi_{E}
\left(\Omega \right)$ satisfies the time independent
Schr\"{o}dinger equation $\hat{h}\left(\Omega \right) \varphi_{E}
\left(\Omega \right)= E\varphi_{E} \left(\Omega \right)$. Given
that $\hat{h}$ is a square root operator, one has to use the
spectral theorem and solve the derived equation
$\hat{h}^{2}\left(\Omega \right) \varphi_{E} \left(\Omega \right)=
E^{2}\varphi_{E} \left(\Omega \right)$ \cite{isham}. The general
solution to this last equation is
\begin{equation}
    \varphi_{\omega} \left(\Omega
\right)= \left[c(\omega)I_{i\omega/2}\left(\sqrt{|2\overline
c|}e^{3\om}\right)+d(\omega)K_{i\omega/2}\left(\sqrt{|2\overline
c|}e^{3\om}\right)\right]
\end{equation}
The functions $I_{i\omega/2}\left(\sqrt{|2\overline
c|}e^{3\om}\right)$ have to be discarded because they diverge in
the classical forbidden zone $\Omega \rightarrow \infty$. The
stationary solutions of the Schr\"{o}dinger equations (\ref{682})
and (\ref{683}) are then of the form
\begin{equation}\label{098}
\Psi^{\omega}_{\pm}(\om,s)=
d(\omega)K_{i\omega/2}\left(\sqrt{|2\overline c|}e^{3\om}\right)
e^{\mp i\omega s}
\end{equation}

It is important to remark that the equation
$\hat{h}^{2}\left(\Omega \right) \varphi_{E} \left(\Omega \right)=
E^{2}\varphi_{E} \left(\Omega \right)$ is the same  that
one obtains from the Wheeler-DeWitt equation associated to the
constraint (\ref{joi}) by separating variables (i.e., by proposing
solutions of the form $\Psi(\om,s)=\varphi \left(\Omega
\right)\phi\left(s \right)$). This means that the equation that
one has to solve \emph{in any case} is a second order
Wheeler-DeWitt equation of the hyperbolic type. The reduction
process changes only the variables in which this equation will be
solved. The only difference is that in the new set of variables it
is clear which are the boundary conditions to be imposed.

\subsection {Relation between solutions of the Schr\"{o}dinger and
the Wheeler-DeWitt
equations}

As we have shown in the preceding examples, there are mainly two
quantization schemes. Being the Wheeler-DeWitt equation an
hyperbolic equation quadratic in all its derivatives, its space of
solutions is twice the space of solutions of the parabolic
Schr\"{o}dinger equation linear in $\frac{\partial}{\partial t}$.
If the reduced Hamiltonian $h$ does not depend on time, the space
of solutions of the Wheeler-DeWitt equation
$\widehat{H}\Psi=0$ will be simply the direct sum of the spaces
of solutions of the Schr\"{o}dinger equations corresponding to
each sheet of the Hamiltonian constraint:
\begin{equation}
Ker \widehat{H}=Ker \widehat{K_{+}} \oplus Ker \widehat{K_{-}},
\end{equation}
where $\widehat{K_{+}}$ and $\widehat{K_{-}}$ are the operators
corresponding to the factors $p_{0}+h$ and $p_{0}-h$ respectively.

It could happen nevertheless that, even if the reduced Hamiltonian
$h$ is time independent, the correspondence between the solutions
of the Wheeler-DeWitt equation and the solutions of the
Schr\"{o}dinger equation is not so direct. For the dilaton
cosmology described by the Hamiltonian constraint (\ref{pop}), the
passage from the Hamiltonian constraint in its original form to
the factorized form (\ref{lse}) is mediated by the canonical
transformation given by
 (\ref{471}). The solutions (\ref{rtu}) of the
Wheeler-DeWitt equation (\ref{rty}) and the solutions of the
Schr\"{o}dinger equations (\ref{682}) and (\ref{683}) are not
expressed in terms of the same variables ($\left(\Omega,
\phi \right)$ and $\left(\Omega, s \right)$ respectively). In
Ref. \cite{cafe01}, this situation was analyzed for the Taub model
and a certain criterium was proposed for fixing boundary
conditions to the Wheeler-DeWitt equation for those cases where we
do not know how to reduce the system. We will now describe such a
proposal.

In Ref. \cite{cafe01} it was proposed that the solutions of the
Wheeler-DeWitt equation (\ref{rty}) can be related with the
solutions of the Schr\"{o}dinger equations (\ref{682}) and
(\ref{683}) by means of a quantum version of the classical
canonical transformation (\ref{471}) used to reduce the system.
Under certain conditions two quantum systems whose Hamiltonians
are canonically equivalent at the classical level, have as quantum
states wave functions which can be related by means of the so
called ``quantum canonical transformations''. This quantum version
of the classical canonical transformations can be understood as a
generalization of the Fourier Transform. It is in fact possible to
consider the Fourier transform as the quantum version of the
classical canonical transformation which interchanges coordinates
and momenta. The generating function of such a canonical
transformation is $F_{1}\left(q, Q \right)=Qq$ and the equations
defining the transformation are
$$p=\frac{\partial F_{1}}{\partial q}=Q ,\ \ \ \ \ \ \ \  P=-
\frac{\partial F_{1}}{\partial Q}=-q.$$
The Fourier transform
\begin{equation}
\Psi \left(q \right)=N\int dp e^{ipq}\Phi \left(p \right)
\end{equation}
can then be rewritten as
\begin{equation}\label{374}
\Psi \left(q \right)=N\int dQ e^{iF_{1} \left(q,Q \right)}\Phi
\left(Q \right),
\end{equation}
The inverse of this transformation could be expressed as
\begin{equation}
\Phi \left(Q \right)=N\int dq
\left|\frac{\partial^{2}F_{1}\left(q, Q \right)}{\partial q
\partial Q} \right| e^{-iF_{1} \left(q,Q \right)}\Psi \left(q \right).
\end{equation}

It is then natural to ask if these expressions remain valid for a
canonical transformation given by a general generating function
$F_{1}\left(q, Q \right)$. If this were the case, one would have a
Generalized Fourier Transform between the quantum representations
associated to systems canonically equivalent at the classical
level. In general this is not the case: certain conditions must be
fulfilled in order to have this kind of Generalized Fourier
Transforms. In \cite{ghan} it was shown that the expression
(\ref{374}) is valid if the following condition is satisfied
\begin{equation}
H_{q}\left(-i\frac{\partial}{\partial q}, q \right)e^{iF_{1}
\left(q,Q \right)}=H_{Q}\left(i\frac{\partial}{\partial Q}, Q
\right)e^{iF_{1} \left(q,Q \right)},
\end{equation}
where certain boundary conditions in the integration limits are
also assumed. If the canonical transformation is defined by means
of other kind of generating function ($F_{2}$, $F_{3}$ or $F_{4}$)
analogous expressions can be used.

In particular, the canonical transformation
(\ref{471}) used to pass from the Hamiltonian constraint
(\ref{pop}) to the factorized Hamiltonian constraint (\ref{lse})
effectively satisfies these requirements. The solutions of the
Schr\"{o}dinger equations (\ref{682}) and (\ref{683}) can then be
related to the solutions of the Wheeler-DeWitt equation
(\ref{rty}) by means of the corresponding Generalized Fourier
Transform.

By means of this formalism we can now analyze the proper boundary
conditions that should be imposed on the space of solutions of the
Wheeler-DeWitt equation (\ref{rty}). Firstly, one should notice
that the dependance on $\Omega$ is the same for both
representations ($\left(\Omega, \phi \right)$ and $\left(\Omega, s
\right)$). This means that one can apply, on the factor depending
on $\Omega$ in the solutions (\ref{rtu}) of the Wheeler-DeWitt
equation, the same boundary conditions that we have previously
imposed on the solutions of the Schr\"{o}dinger equations. In
fact, given our choice of the physical clock as a function of both
$\phi$ and $p_{\phi}$ (see (\ref{471})), the variable $\Omega$ is
an authentical dynamical variable. In this way one can discard in
the solution (\ref{rtu}) the functions
$I_{i\omega/2}\left(\sqrt{|2\overline c|}e^{3\om}\right)$ because
they diverge in the classical forbidden zone $\Omega \rightarrow
\infty$. Next we have to impose boundary conditions on the factor
which depends on $\phi$ in (\ref{rtu}), i.e., we have to decide if
we will discard the functions
$I_{i\omega}\left({|\lambda|}e^{-\phi}\right)$ or the functions
$K_{i\omega}\left({|\lambda|}e^{-\phi}\right)$. The criterium that
we will apply is that the physical solutions will be those whose
transformed functions (via the generalized Fourier transform)
coincide with the factors $e^{-i\omega s}$ or $e^{i\omega s}$ in
the solutions $\Psi\pm$ (\ref{098}) of the Schr\"{o}dinger
equations (\ref{682}) and (\ref{683}) for the clock choices $t=s$
or $t=-s$ respectively. One could suppose that the correct
solutions will be those which go to zero in the classical
forbidden zone $\phi \rightarrow -\infty$, i.e., the functions
$K_{i\omega}\left({|\lambda|}e^{-\phi}\right)$. But if we
transform this functions we have to conclude that they have to be
discarded given that they correspond to a linear combination of
the form $ae^{i\omega s}+be^{-i\omega s}$. If on the contrary we
apply the generalized Fourier transform to the functions $I_{\pm
i\omega}\left({|\lambda|}e^{-\phi}\right)$ we obtain the
correspondence $I_{\pm i\omega}\left({|\lambda|}e^{-\phi}\right)
\leftrightarrow e^{\mp i\omega s}$. The functions
$I_{i\omega}\left({|\lambda|}e^{-\phi}\right)$ and
$I_{-i\omega}\left({|\lambda|}e^{-\phi}\right)$ do represent then
the positive energy states corresponding to the clock choices
$t=s$ and $t=-s$ respectively. If we make the choice $t=s$ the
solution to the Wheeler-DeWitt equation (\ref{rty}) is then
\begin{equation}\label{lkj}
\Psi_\omega(\om,\phi) = \tilde{a}\left(\omega \right)
I_{i\omega}\left({|\lambda|}e^{-\phi}\right)K_{i\omega/2}\left(\sqrt{|2\overline
c|}e^{3\om}\right)
\end{equation}

The fact that the correct solutions are not those which go to zero
in the classical forbidden zone $\phi \rightarrow -\infty$ does
not pose a problem given that the variable $\phi$ is not a
dynamical variable, but the variable associated with the physical
clock that we have chosen. This fact constitutes an important
difference with the immediate result that we would obtain by
following Ref. \cite{moncrief}.

It could be argued that the real problem is to find proper
boundary conditions for the Wheeler-DeWitt equation when one does
not know how to reduce the system. If one has a reduced system
with the corresponding Schr\"{o}dinger equation as in the Taub
case, it is no more necessary to go back to the Wheeler-DeWitt
``representation''. However, given that for the Taub model both
spaces of solutions are known, its analysis is of great utility in
order to propose general boundary conditions to the Wheeler-DeWitt
equation, even for those cases for which one does not know how to
separate a physical clock. The main requirement to impose on the
general boundary conditions that we are looking for is that, for
those cases for which one knows how to reduce the model (as the
Taub universe), they have to select the same quantum states
imposed by the quantization of the reduced system. In Ref.
\cite{cafe01} certain steps were given in this direction. We will
now describe these results.

Let us suppose as a first simplification that we have the
following scaled Hamiltonian constraint
\begin{equation}\label{uiy}
H=p_{0}^{2}+V\left(q^{0}\right)-h^{2}\left(q^{\mu}, p_{\mu}\right)
\end{equation}
with $V\left(q^{0}\right)>0$. In other words, let us suppose that
there is not a non-minimal coupling between a certain variable
$q^{0}$ and the rest of the canonical variables $q^{\mu}$. Given
this form of the Hamiltonian constraint, the solutions of the
corresponding Wheeler-DeWitt equation will have the form $\Psi
\left(q^{0}, q^{\mu}\right)=\Theta\left(q^{0}
\right)\Phi\left(q^{\mu}\right)$. If the potential
$V\left(q^{0}\right)$ were identically zero, the variable $q^{0}$
would be a physical clock. We could suppose that the true physical
clock will be a certain function of $q^{0}$ which, in the region
where $V\left(q^{0}\right)\rightarrow 0$, coincides with $q^{0}$.
We can expect then that the solutions $\Psi \left(q^{0},
q^{\mu}\right)$ of the Wheeler-DeWitt equation corresponding to
(\ref{uiy}) would tend, in the region where
$V\left(q^{0}\right)\rightarrow 0$, to wave functions of the form
$e^{-iEq^{0}}\Phi\left(q^{\mu}\right)$. The boundary conditions to
be imposed on the space of solutions of the Wheeler-DeWitt
equation is that the physical solutions will be those functions
whose asymptotic expressions in the region where
$V\left(q^{0}\right)\rightarrow 0$ are of the form
$e^{-iEq^{0}}\Phi\left(q^{\mu}\right)$ (with
$\Phi\left(q^{\mu}\right)$ going to zero in the classical
forbidden zone associated to $q^{\mu}$). If we do not know which
is the physical clock for reducing the system, but we know that
there is a variable $q^{0}$ which is a physical clock \emph{in a
certain limited region} with a time independent reduced
Hamiltonian $h$, then we can imposed as boundary conditions that
the physical solutions have to tend in that region to functions
with a factor $e^{-iEq^{0}}$.

If we now apply this criterium to the solutions of the
Wheeler-DeWitt equation (\ref{rty}) corresponding to the Taub
model, we will select the correct quantum states, i.e. the quantum
states (\ref{lkj}) selected by the reduction of the model. The
Hamiltonian constraint (\ref{popy}) of the Taub model has in fact
the form (\ref{uiy}). In the region where the potential term
$\lambda^{2}e^{-2\phi}$ goes to zero, i.e., in the region $\phi
\rightarrow \infty$, the variable $\phi$ is a physical clock.
Following the proposed criterium, we have to select those
functions which tend to functions with a factor $e^{-iE \phi}$ in
the region $\phi \rightarrow \infty$. If we consider the
asymptotic expressions of the functions
$K_{i\omega}\left({|\lambda|}e^{-\phi}\right)$ and $I_{\pm
i\omega}\left({|\lambda|}e^{-\phi}\right)$ in that limit, we will
find that the former tend to a linear combination of the form
$ae^{iE \phi}+be^{-iE \phi}$, while only the latter tend to
functions of the form $e^{\pm iE \phi}$.

\subsection{Time dependent reduced Hamiltonians}

A serious problem for the understanding of these two quantization
formalisms and its relations appears when the reduced Hamiltonian
$h$ is time dependent. While the constraints $H=p_{0}^{2}-h^{2}=0$
and $H=\left(p_{0}+h\right)\left(p_{0}-h\right)=0$ are classically
equivalent, at the quantum level this equivalence is no more
fulfilled if the reduced Hamiltonian $h$ depends on the variable
chosen as physical clock. In fact a wave function in the kernel of
the operators $\widehat{K}_{+}$ and $\widehat{K}_{-}$
corresponding to the factors $\left(p_{0}+h\right)$ and
$\left(p_{0}-h\right)$ respectively (i.e., a solution of the
Schr\"{o}dinger equation), is not necessarily annihilated by the
operator $\widehat{H}$ associated with the Wheeler-DeWitt
equation. If the reduced Hamiltonian $h$ is time dependent the
product of the Schr\"{o}dinger operators $\widehat{K}_{+}$ and
$\widehat{K}_{-}$ is not equal to the Wheeler-DeWitt operator
$\widehat{H}$. The two possible products of the Schr\"{o}dinger
operators are
\begin{eqnarray}
\widehat{K}_{\pm}\widehat{K}_{\mp}=-\frac{\partial^{2}}{\partial
t^{2}} -\hat{h}^{2}\mp \left[-i\frac{\partial}{\partial t},
\hat{h} \right]
\end{eqnarray}
If $h \neq h(t)$ (i.e., $\left[p_{t}, h \right]=0$), then
$\widehat{K}_{+}\widehat{K}_{-}=\widehat{K}_{-}\widehat{K}_{+}=\widehat{H}=-\frac{\partial^{2}}{\partial
t^{2}} -\hat{h}^{2}$. If $h = h(t)$, then the Wheeler-DeWitt
operator $\hat H$ is equal to
\begin{eqnarray}
\widehat{H}=\frac{1}{2}\left(\widehat{K}_{+} \widehat{K}_{-} +
\widehat{K}_{-} \widehat{K}_{+} \right).
\end{eqnarray}
It is then clear that the solutions to the Schr\"{o}dinger
equations $\widehat{K}_{\pm}\Psi_{\pm}=0$ are not necessarily
solutions of the Wheeler-DeWitt equation $\widehat{H} \Psi=0$ when
the reduced Hamiltonian $h$ is time dependent \cite{isham}. As it
was explained in \cite{isham}, if the reduced Hamiltonian $h$ is
time dependent the solution to the Schr\"{o}dinger equation $i
\frac{\partial \Psi \left(x, t \right)}{\partial t}=\hat{h}
\left(x, p, t \right)\Psi\left(x, t \right)$ takes the form
\begin{equation}\label{gol}
\Psi\left(x, t\right)=
T\left[e^{-i\int_{t_{0}}^{t}\hat{h}\left(t'\right)dt'}
\right]\Psi\left(x, t_{0} \right),
\end{equation}
where $T$ is the time-ordering operator. If the condition
\begin{equation}\label{sde}
\left[h\left(t\right), h\left(t'\right) \right]=0
\end{equation}
is satisfied, the expression (\ref{gol}) gives
\begin{equation}
\Psi\left(x, t\right)=
e^{-i\int_{t_{0}}^{t}\hat{h}\left(t'\right)dt'} \Psi\left(x, t_{0}
\right).
\end{equation}

The condition (\ref{sde}) implies also that there exists a
conserved complete set of eigenstates of the Hamiltonian operator
$\hat{h}\left(t \right)$, i.e., a set of basis states which are
eigenstates of $\hat{h}$ at all times. If $\Psi_{E}\left(x
\right)$ is an eigenstate of $\hat{h}\left(t_{0} \right)$, then
$\Psi_{E}\left(x \right)$ will be an eigenstate of $\hat{h}\left(t
\right)$ for all times, i.e., there will be a function $E\left(t
\right)$ such that
\begin{equation}
\hat{h}\left(t \right) \Psi_{E}\left(x \right)= E\left(t
\right)\Psi_{E}\left(x \right).
\end{equation}
The time evolution of such a state is given by
\begin{equation}
\Psi_{E}\left(x , t \right)= e^{-i \int_{t_{0}}^{t}E
\left(t'\right)dt'}\Psi_{E}\left(x , t_{0} \right).
\end{equation}

An example of a system with a time dependent reduced Hamiltonian
$h(t)$ is provided by the dilatonic cosmological model
corresponding to the Hamiltonian constraint (\ref{927}) for
$\zeta>0$. The factorized form of this constraint is
\begin{equation}
H=\left(-p_x+\sqrt{ p_y^2+\zeta e^{2x}}\right)\left( p_x+\sqrt{
p_y^2+\zeta e^{2x}}\right)= 0,
\end{equation}
being the potential time dependent for $t=\pm x$. Therefore,
though at the classical level this product is equivalent to the
constraint (\ref{927}), in its quantum version both constraints
differ in terms associated to commutators between $\hat{p}_x$ and
the potential $\zeta e^{2x}$. The  general form of these
commutators is $\left[\sqrt{\sum(\hat{ p_\mu})^2 +V(\hat{
q^i})},\hat{ p}_0\right]$ (where $\mu\neq 0$, and $V$  stands for
the potential in the scaled Hamiltonian constraint $H$).
Hence, depending on which of the two classically equivalent
constraints we start from, we obtain different quantum theories.
Observe that this problem appears in the case for which the
Wheeler-DeWitt equation leads to a result in which the usual
identification of positive and negative-energy solutions is not
apparent, at least in the standard form of exponentials of the
form $e^{i\omega t}$. In this sense, recall the difference between
the solutions (\ref{tt}) and (\ref{ttt}).

The central obstruction for the existence of a trivial
correspondence between the Wheeler-DeWitt and Schr\"{o}dinger
solutions for minisuperspaces is then a constraint with a
time-dependent potential. For the class of models of Section 3.2,
a coordinate choice avoiding the decision between inequivalent
quantum theories can be introduced \cite{si03}. Consider the
constraint (\ref{613}) and define
\begin{equation} u = \alpha
e^{{1\over 2}(a q^1+b q^2)}\cosh
\left({b q^1+a q^2\over 2}\right),\ \ \ \ \   v =\alpha
e^{{1\over 2}(aq^1+b q^2)}\sinh
\left({b q^1+a q^2\over 2}\right),
\end{equation}
with $\alpha=\sqrt{|A|}$. These coordinates allow to write the
scaled constraint in the equivalent form \be H= -p_u^2+p_v^2+\eta
m^2=0, \ee with $\eta=sgn (A)$ and $m^2=4/|a^2-b^2|$. Given that
the commutators do not appear now, the Wheeler-DeWitt equation is
equivalent to the corresponding Schr\"{o}dinger equations, being
the physical clock the coordinates $u$ or $v$ depending on $\eta$: for $\eta=1$ we have $p_u\neq 0$ and $t=\pm u$, while for $\eta=-1$ we have $p_v\neq 0$ and $t=\pm v$.

Of course, such a  solution can be applied only for a limited
number of minisuperspace models. We believe that the case of a
time dependent reduced Hamiltonian $h=h(t)$ is the general case.
In fact it seems highly improbable that one could find a degree of
freedom for playing the role of a physical clock which would not
be coupled to the others degrees of freedom (besides the minimal
coupling given by the Hamiltonian constraint). In other words, we
would not expect to find a ``free clock'', i.e. a degree of
freedom without a non-minimal coupling with the others degrees of
freedom. The reduced system will be then in general an open
system, i.e., a system which interacts with the clock
\cite{desi99a}. If the reduced Hamiltonian $h$ is time dependent,
the energy of the reduced system will change. This fact could seem
bizarre if one takes into account that in quantum cosmology we are
considering the whole universe. But in that case the physical
clock will have its own energy, being the change in the energy of
the reduced system a consequence of its interaction with the clock
(which is nothing but a particular interacting degree of freedom).
This change in the ``reduced'' energy of the system will
correspond then to a variation in the momentum of the physical
clock ($p_{t}=-h$): if the clock changes its rate of evolution,
the total energy of the others degrees of freedom will change.

\section{General aspects of the problem of time}

\subsection{Parametrized system formalism}

A general framework for understanding the meaning of the
Hamiltonian constraint $H=0$ is provided by the formalism of
parametrized systems (for details, see \cite{hete92}; for recent developments,
see for example \cite{fesi97,tkach,cafe01,ccf,bale01,vara04,bale04}).
In this formalism the ``real'' time $t$ is
added to the canonical variables of an ordinary dynamical system,
being the increased set of variables left as a function of a
physically irrelevant parameter $\tau$. The ordinary action of a
dynamical system
\begin{eqnarray}
S\left[ q^{\mu },p_{\mu }\right] =\int_{t_{1}}^{t_{2}}p_{\mu
}dq^{\mu }-h\left( q^{\mu },p_{\mu }\right) dt,\ \ \mu =1,...,n
 \label{719}
\end{eqnarray}
can then be converted into a parametrized action by means of the
definition of a new pair of canonical variables $\left\{
q^{0}=t,p_{0}=-h \right\}$ associated to the time $t$ and the
Hamiltonian $h$. The extended set of variables are left as
functions of the physically irrelevant parameter $\tau $. \ The
set $\left\{ q^{0},q^{\mu },p_{0},p_{\mu }\right\}$ can be varied
independently, provided that the definition of $p_{0}$ is
incorporated into the action as a constraint
\begin{equation}
H=p_{0}+h=0,  \label{po}
\end{equation}
with the corresponding Lagrange multiplier $N$ (lapse
function). In this way one obtains the following action
\begin{equation}
S\left[ q^{i}\left( \tau \right) ,p_{i}\left( \tau \right)
,N\left( \tau \right) \right] =\int_{t_{1}}^{t_{2}}\left(
p_{i}\frac{dq^{i}}{d\tau } -NH\right) d\tau,\ \ i =0,...,n
\label{pl}
\end{equation}

The presence of the Lagrange multiplier $N\left( \tau \right) $
means that the dynamics remains ambiguous in the irrelevant
parameter $\tau $ (one could say that it has no sense to speak
about dynamics until the hidden time is recovered). In the
extended configuration space $\left\{q^{0}=t,q^{\mu} \right\}$ the
solutions of the equations of motions are static ``curves''
without a preferred parametrization. The theory is then invariant
under reparametrizations of the physical irrelevant parameter
$\tau$. We could say that it has no sense to speak about the speed
of the motion of the system through its extended configuration
space. These curves have nevertheless a preferred sense: they
unfold in the increasing directions of $q^{0}=t$.

In this way any dynamical system can be formulated in the
framework of parame\-trized systems. In the process of
parametrization, the real time is ``disguised" as a dynamical
variable. This disguised time can be nevertheless easily
recognized due to the linearity of its conjugated momentum in the
constraint (\ref{po}).

It has been intended to use the formalism of parametrized system
as a model for understanding the canonical structure of General
Relativity. Having the theory a Hamiltonian constraint of the form
$H=0$, one could suppose that General Relativity is an ordinary
dynamical theory whose action is presented in a parametrized form.
If that were the case, one could try to reduce the system by
finding the disguised ``real" time and formulating the theory as
an ordinary dynamical system of the form (\ref{719}), whose
evolution takes place in that ``real'' time variable. There has
been many proposals for that hidden time variable but none of them
could circumvent the different problems which appear in the
reduction process \cite{ku81, ku92}. Besides the lack of a
universal ``real" time variable, there is still another important
objection against this interpretation of the canonical structure
of General Relativity: the constraint (\ref{po}) is linear in the
momentum $p_{0}$, while the Hamiltonian constraint of General
Relativity is quadratic in all its momenta.

In \cite{ccf} a different interpretation was proposed in order to
circumvent this problem. Here we will follow this new conceptual
framework. We will consider that the supposition of a privileged
``real" time hidden among the canonical variables is an unfounded
supposition. We consider that one of the most fundamental
properties of General Relativity is that its solutions do not
represent an \emph{evolution in time} of certain dynamical
variables, but that it is a theory which selects certain relative
(not dynamical) configurations of its canonical variables which,
under certain conditions, can be considered as dynamical
evolutions if proper \emph{physical clocks} are selected. In fact
we can never observe the evolution of the degrees of freedom along
a ``newtonian'' time flow like $q^{1}\left( t\right)$ and
$q^{2}\left( t\right)$ but rather the evolution of certain
variables relative to the change of another variable, i.e.,
something like $q^{2}\left(q^{1}\right)$. Following this
interpretation, there would not be such a thing as Time, but only
physical degrees of freedom which, under certain conditions, can
be used as physical clocks. In this relational approach we cannot
say that reducing the system means to find the hidden ``real''
time: in order to reduce the model we have only to select a
monotonously increasing canonical variable as a physical clock. It
is thus only possible to speak about physical clocks, i.e.,
degrees of freedom which, under certain conditions, can play the
role of evolution parameters for the others degrees of freedom. In
that case it is not important if the physical clock is a
geometrical degree of freedom or a degree of freedom associated to
the matter fields (see, for example, \cite{ku95}); in fact, among
the examples of sections 3.2 and 3.3, we identified models with
the dilaton field playing the role of physical clock. On the
contrary, if we were looking for the ``real'' time, we could
expect that this hidden time would be a geometrical degree of
freedom.

The works  \cite{htv92,fesi97,si98,si99,desi99b} have shown that
the general covariance of General Relativity is not so different
from the gauge
symmetry of an ordinary gauge theory (see Appendix B). In fact, we
could consider General Relativity as a particular case of a gauge
theory. In general, in gauge theories the gauge fixing is not
interpreted as the discovery of the ``real'' degrees of freedom,
but as a particular choice without any ``god given'' privilege. In
the framework  of gravitation the election of a physical clock
acts in fact as a partial gauge fixing of the theory (election of
the Lagrange multiplier $N\left( \tau \right)$). If we follow the
ordinary interpretation of gauge theories this election should not
be interpreted as the discovery of the real hidden time, but as a
choice of a particular reference system for measuring the
evolution of the system.

We could now ask which is the proper formalism for implementing
such an interpretation. In \cite{ccf} a certain modification of
the parametrized formalism was proposed in order to explore the
consequences of this new conceptual framework.

The first point that one should take into account is that, as
there is not a privileged time variable $t$ and a privileged
conjugated momentum $p_{t}=-h$, all the momenta must appear on an
equal footing in the Hamiltonian constraint (as effectively
happens in General Relativity). This means that there should not
be a preferred momentum appearing linearly as in (\ref{po}). It is
then a consequence of the new interpretation the necessity of
finding a formalism where all the momenta appear quadratically in
the Hamiltonian constraint.

There is still a second argument for implementing a formalism of
parametrized systems with a constraint quadratic in all its
momenta. The solutions of the theory are static trajectories in a
certain configuration space. In order to describe these static
trajectories as dynamical evolutions, we have to choose a certain
variable to play the role of time. The main difference with the
original formalism of parametrized systems is that these
trajectories do not carry a preferred sense of evolution. We can
choose to unfold them in both directions. This ambiguity can be
cast into the parametrized system formalism by means of a modified
Hamiltonian constraint. If there is not a privileged time the
solutions are necessarily static trajectories, i.e., relative
configurations among the different variables, for example
$q^{1}\left(q^{0}\right) $. If one wants now to select a physical
clock, for example $q^{0}$ (we are supposing that $q^{0}$ is a
monotonously increasing function along the trajectory) there is
still an ambiguity: one still has to choose in which direction the
trajectory will be unfold. This means that one can choose
$t=q^{0}$ or $t=-q^{0}.$ The static trajectory does not privilege
any direction and then both kinds of solutions must appear in the
reduced formalism. These two options correspond to the constraints
$K_{+}=p_{0}+h=0$ and $K_{-}=p_{0}-h=0$. Both possibilities can be
incorporated into the formalism if we use a Hamiltonian constraint
of the form $H=\left(p_{0}+h \right) \left(p_{0}-h
\right)=p_0^2-h^2=0$. In order to make this factorization the
original Hamiltonian constraint must be quadratic in the momentum
conjugated to $q^{0}$. The existence of two sheets in the
Hamiltonian constraint, far from being an unnecessary redundance,
acquires in this way a precise meaning related to the necessity of
having a parametrized formalism which does not privilege any sense
of evolution. Each sheet of the Hamiltonian constraint $H=0$ is
then associated to each possible choice of the direction in which
the trajectory can be unfold. The resulting action is
\begin{equation}
S\left[ q^{i}\left( \tau \right) ,p_{i}\left( \tau \right)
,N\left( \tau \right) \right] =\int p_{\mu }dq^{\mu
}+p_{0}dq^{0}-N \left(p_0^2-h^2 \right) d\tau
\end{equation}

The constraint $H=0$\ is fulfilled if one of the factors vanishes
on the constraint surface. To choose which factor is null is
equivalent to choose which direction of $q^{0}$ is the increasing
direction of time. This system can then be reduced by the clock
choices $t=q^{0}$ or $t=-q^{0}$, corresponding these choices to
the sheets $p_{0}+h=0$ and $p_{0}-h=0$ respectively. If $t=q^{0}$,
then $p_{t}=p_{0}=-h$; if $t=-q^{0}$, then $p_{t}=-p_{0}=-h$. In
both cases $p_{t}=-h$ with $h>0$. Having fixed a factor to zero in
the Hamiltonian constraint the other one will have, on the
constraint surface, a definite sign, so being possible to rescale
the Hamiltonian by this factor. If for example we fix
$K_{+}=p_{0}+h=0$ ($t=q^{0}$ as physical clock), we will have
$p_{0}=-h <0$ and then the other factor will be
$K_{-}=p_{0}-h=-2h<0$.

In this way, both the necessity of having a Hamiltonian constraint
where all the momenta appear on an equal footing (quadratically)
and the necessity of having the possibility of choosing the
direction of unfolding of any static trajectory in both
directions, conduce naturally to a modification of the
parametrized formalism. We believe that this \emph{quadratic
parametrized formalism} is the proper formalism for understanding
the canonical structure of General Relativity.

\subsection{Motion-reversal and clock-reversal transformations}

Without being precise, one could say each choice in the sense of
evolution of the physical clock ($t=q^{0}$ or $t=-q^{0}$)
corresponds to a kind of ``time reversal'' of the other one. But by
choosing a physical clock, i.e., by choosing a sheet of the
Hamiltonian constraint $H=\left(p_{0}+h \right) \left(p_{0}-h
\right)=0$, one obtains an ordinary dynamical system. It is well
known that an ordinary classical or quantum system possess a
certain symmetry under ``time reversals''. If each sheet posses a
symmetry under ``time reversals", the passage from one sheet to the
other one can not be identified with the same time reversal
operation. In \cite{ccf} this situation was clarified. The
result of this analysis is that there are two kind of ``time
reversals'' operations which have to be carefully differentiated.

The first one is the ordinary ``time reversal'' operation of
classical and quantum mechanics. In fact this operation does not
correspond to an inversion in the direction of time, but to an
inverted movement which unfolds in the same direction of time than
the original solution. In \cite{ccf} this operation was
called \emph{motion-reversal transformation}. Given a classical
trajectory $\left\{ q\left( q_{0},p_{0},t_{0},t\right) ,p\left(
q_{0},p_{0},t_{0},t\right) \right\} $ which unfolds from
$\left\{ q_{0},p_{0}\right\}$ at time $t_{0}$ to
$\left\{q_{f},p_{f}\right\}$ at time $t_{f}$, there exists a
related trajectory which is also a solution of the same Hamilton
equations. This inverted trajectory is
\begin{eqnarray}
q^{mr}\left( q_{0}^{mr}=q_{f},p_{0}^{mr}=-p_{f},t_{0,}t\right)
=q\left( q_{f},-p_{f},t_{0,}t\right),  \label{oi} \\ p^{mr}\left(
q_{0}^{mr}=q_{f},p_{0}^{mr}=-p_{f},t_{0,}t\right) =p\left(
q_{f},-p_{f},t_{0,}t\right), \nonumber
\end{eqnarray}
and exists provided that the Hamiltonian $h$ is quadratic in $p$
and does not depend on $t$. This motion-reversed solution $\left\{
q^{mr},p^{mr}\right\}$ unfolds from $\left\{q_{f},p_{f}\right\}$
at time $t_{0}$ to $\left\{ q_{0},p_{0}\right\}$ at time $t_{f}$.
It is important to emphasize that this motion-reversed solution is
a solution of the \emph{same} Hamilton equations. This solution
starts at the same time $t_{0}$ than the original one and unfolds
in the same direction of time, but with initial conditions which
have been inverted with respect to the original trajectory: the
new trajectory starts with an inverted momentum from the point
where the original one ends. This operation does not correspond to
the operation of ``playing the film backwards''. In order to see
this it is necessary to realize that, being the clock a dynamical
variable, it has to be included in the hypothetical film. If we
play the film backwards, we will see the clock running backwards.
However, we have just said that the motion-reversal transformation
of a given solution unfolds in the same direction of time, i.e.,
the clock must continue to run forward. In other words, the
motion-reversed solution corresponds, not to play the same film
backwards but to play \emph{another} film in which all behaves as
running backwards, but the clock is still running forward. The
motion-reversal transformation reverses then all variables,
\emph{but the variable identified with physical clock}. In the
extended configuration space which includes the physical clock the
result of a motion-reversal transformation it is not the same
curve (i.e., the same film) unfolded in the opposite direction,
but \emph{another} curve which unfolds in the same direction of
time.

On the other hand we have a symmetry associated to the passage
from one sheet of the Hamiltonian constraint $H=\left(p_{0}+h
\right) \left(p_{0}-h \right)=0$ to the other one. This operation
involves \emph{also} a change in the direction of unfolding of the
variable used as physical clock. This operation was called in
\cite{ccf} \emph{clock-reversal transformation}. The
clock-reversed solution will be a solution of the Hamilton
equations corresponding to the other sheet of the Hamiltonian
constraint. This operation does correspond now to the operation of
``playing the film backwards''. In the extended configuration
space which includes the physical clock, the graph of the solution
obtained by means of this operation coincides with the original
one, being the only difference the sense of evolution. In other
words, it is now the same film (i.e., the same curve in the
extended configuration space) played in the reversed sense.

Given a solution of the equations of motion in one of the sheets
of the Hamiltonian constraint, it is then possible to construct
three others related solutions: the motion-reversal (in the same
sheet as the original one), the clock-reversal (in the other
sheet), and the motion reversal of the clock-reversed solution (in
the other sheet). Summarizing, we have two solutions in each sheet.

The main difference between the linear formalism for parametrized
systems and the quadratic one is that the latter is invariant
under clock-reversal transformations. In general, any dynamical
trajectory (which carries then a \emph{fixed} sense of evolution)
can be considered as a static one by considering the clock in an
extended configuration space. Inversely, any static trajectory
(without a \emph{fixed} sense of evolution) can be
``temporalized'' by considering a monotonously increasing variable
along the trajectory as a physical clock. We could say that if the
linear formalism is the proper formalism for considering the
solutions of a dynamical system as static trajectories in an
extended configuration space, the quadratic formalism is the
proper one for expressing the static solutions as dynamical ones.
In the first case the dynamical solution that one wants to render
static has its own sense of evolution. It is then unnecessary to
consider a Hamiltonian constraint with a two sheet structure in
the process of parametrization. On the contrary, in the second
case there is not a preferred sense of evolution for the
``temporalization'' of the static trajectory. Given that all the
canonical variables are true degrees of freedom, there is not a
``true'' time disguised among them. This absence of a true time
means that there is not a preferred sense of evolution for
unfolding the trajectory. The Hamiltonian constraint must include
then the two possible senses of evolution, which means that it
must be quadratic in all its momenta.

In a quantum mechanical context these two operations assume very
simple forms. Given a particular solution $\Psi_{+} \left(q^{\mu},
q^{0}\right)$ to the Schr\"{o}dinger equation
\begin{equation}\label{ghj}
i \frac{\partial}{\partial q^{0}}\Psi_{+} \left(q^{\mu},
q^{0}\right)=\hat{h} \Psi_{+} \left(q^{\mu}, q^{0} \right)
\end{equation}
corresponding to the sheet $p_{0}+h=0$ ($t=q^{0}$), its
motion-reversed solution is given by
\begin{equation}
\Psi^{mr}_{+} \left(q^{\mu}, q^{0} \right) =T\Psi_{+}
\left(q^{\mu}, -q^{0}\right)
 \label{qw}
\end{equation}
where $T$ is an antiunitary operator which, in coordinate
representation, is equal to the complex conjugation operator
$T\Psi \left( q\right) =\Psi ^{\star }\left( q\right)$. We could
say that if the $T$ operation changes the sense of all variables
$q^{i}=\left\{q^{\mu}, q^{0} \right\}$, the substitution
$q^{0}\rightarrow -q^{0}$ cancels this change \emph{only} for the
variable $q^{0}$. It is important to remark that this motion
reversed solution $\Psi^{mr}_{+}$ is a solution of the same
Schr\"{o}dinger equation (\ref{ghj}). On the contrary, the quantum
version of the clock-reversal transformation is performed
\emph{only} by the action of the antiunitary operator $T$ (now we
want to change \emph{also} the sense of $q^{0}$):
\begin{equation}
\Psi^{cr}_{-} \left(q^{\mu}, q^{0} \right) =T \Psi_{+} \left(
q^{\mu}, q^{0} \right).
\end{equation}
This clock reversed solution $\Psi^{cr}_{-}$ is not a solution of
the Schr\"{o}dinger equation (\ref{ghj}) but a solution of the
Schr\"{o}dinger equation corresponding to the other sheet of the
Hamiltonian constraint ($p_{0}-h=0$, $t=-q^{0}$):
\begin{equation}\label{dfr}
-i \frac{\partial}{\partial q^{0}}\Psi_{-} \left(q^{\mu},
q^{0}\right)=\hat{h} \Psi_{-} \left(q^{\mu}, q^{0} \right).
\end{equation}

It is important to remark, contrary to what is usually believed,
that each sheet of the Hamiltonian constraint ($p+h=0$ and
$p-h=0$) corresponds to positive energies solutions. What changes
from one sheet to the other one is the sign of time, not the sign
of the energy. Each sheet corresponds to each possible choice in
the direction in which the static trajectory can be unfolded,
being in both cases positive energy solutions. In fact, the
stationary solutions to the Schr\"{o}dinger equation (\ref{ghj})
are
\begin{equation}
\Psi^{E}_{+} \left(q^{\mu }, q^{0} \right) =e^{-iEq^{0}} \varphi
\left( q_{\mu }\right) =e^{-iEt}\varphi \left(q_{\mu }\right),
\end{equation}
while the stationary solutions to the Schr\"{o}dinger equation
(\ref{dfr}) are
\begin{equation}
\Psi ^{E}_{-}\left(q^{\mu }, q^{0} \right) =e^{iEq^{0}} \varphi
^{*}\left( q_{\mu }\right) =e^{-iEt}\varphi ^{*}\left( q_{\mu
}\right),
\end{equation}
which shows that both sets of solutions are positive energy
solutions for the two possible choices of the physical clock
($t=q^{0}$ or $t=-q^{0}$).

Following this new interpretative framework, the boundary
conditions for the space of solutions of the Wheeler--DeWitt
equation can be interpreted as the \emph{symmetry breaking} of the
clock-reversal invariance of a theory described by a Hamiltonian
constraint with a two sheet structure. The boundary conditions
proposed in Section 3.4 separate the space of solutions in those
quantum states going forward in the time $t=q^{0}$ from those
quantum states going forward in the time $t=-q^{0}$, being these
two subspaces related by the clock-reversal operator $T$. The fact
that the Wheeler-DeWitt operator $H$ is a real operator has as a
consequence that, given a certain solution $\Psi$, the associated
function $\Psi^{*}$ will be also a solution, being these solutions
linearly independent. The space of solutions $S$ of the
Wheeler--DeWitt equation can then be decomposed as the direct sum
$S=C \oplus C^{*}$. In the case of the solution (\ref{rtu}) to the
Wheeler--DeWitt equation (\ref{rty}), the factor depending on
$\phi$ could be expressed equivalently as a linear combination of
the modified Bessel functions $K_{\nu}$ and $I_{\nu}$ or as a
linear combination of the modified Bessel functions $I_{\nu}$ and
$I_{- \nu}$. But while the functions $K_{\nu}$ and $I_{\nu}$ are
not complex conjugated of each other, the functions $I_{\pm \nu}$
do satisfy the property $I_{\nu}=I_{- \nu}^{*}$. As a result, the
decompositions $S=K_{\nu}\oplus I_{\nu}$ and $S=I_{\nu} \oplus
I_{-\nu}$ are not equivalent, being the latter the correct
decomposition for imposing the proposed boundary conditions. If
one can find a decomposition of the space of solutions of the form
$S=C \oplus C^{*}$, then the problem of imposing boundary
conditions on this space of solutions is solved.

An important difference between the clock-reversal and the
motion-reversal transformations is that a theory with a time
dependent reduced Hamiltonian $h \left(t \right)$, even if it is
not symmetric under a motion-reversal transformation, it is still
symmetric under a clock-reversal transformation. This means that
even if it is not possible to reduce the system by means of a time
independent reduced Hamiltonian $h$, the space of solutions of the
Wheeler-DeWitt equation must continue to be a direct sum of two
subspaces related by the antiunitary operator $T$.

\section{Conclusion}

The quantization of string cosmologies is of particular interest because they allow for new points of view about the earliest stages of the universe \cite{gave,ga00}. In the present work we have focused our attention in the formal aspects of the problem within the minisuperspace approximation.  We shall now enumerate the central points of the whole discussion.
As it is well known from quantum cosmology, the most important
fact for the comprehension of the canonical structure and
quantization of dilaton cosmological models is the presence of the
so called \emph{Hamiltonian constraint}. As it was explained in
these notes, a theory with this kind of constraint lacks of a
definition of a ``true'' time variable. A theory with such a
canonical structure can not be directly interpreted in terms of an
evolution in time of its degrees of freedom. In order to
``temporalize'' the theory it is necessary to select a
\emph{physical clock}, i.e., a monotonously increasing variable
which could serve as a time parameter for measuring the evolution
of the rest of the canonical variables. The notion of a physical
clock was not interpreted in terms of a recovered ``real'' hidden
time, but in terms of a particular degree of freedom which
satisfies certain properties which let us use it as a globally
well defined clock (it is then not necessary to search for a
physical clock \emph{only} among the geometrical degrees of
freedom). We have analyzed the consequences of this interpretation
on the so called \emph{parametrized formalism}, which is usually
used for understanding the canonical structure of a theory with a
Hamiltonian constraint \emph{linear} in one of its momenta. It was
then analyzed a modification of this parametrized formalism in
order to better describe the \emph{quadratic} Hamiltonian
constraint of the gravitational models that we are studying.

If a physical clock can be separated, then the theory can be
quantized in a straightforward manner by means of the parabolic
\sch\ equations associated to each sheet of the Hamiltonian
constraint. If one does not know how to separate a physical clock,
one can still quantize the theory by following the so called Dirac
method for quantizing constrained theories. Following this method,
the physical quantum states have to satisfy the operator version
of the quadratic Hamiltonian constraint (\wdw\ equation). The
problem of this method is that, lacking of a definition of time,
we do not know neither how to interpret the corresponding
solutions in terms of a conserved positive-definite probabilities,
nor how to impose boundary conditions on the space of solutions of
the \wdw\ equation. In these notes we have studied certain models
which can be quantized by following both methods (\sch\ and \wdw\
quantizations). The important notion of a \emph{quantum
canonically transformation} was introduced in order to relate the
solutions obtained by both methods. This models are of great
utility in order to propose general boundary conditions for those
cases for which we do not know how to reduce the model by
selecting a physical clock.

Finally we have discussed the temporal symmetries which are  characteristic
of the theories with a quadratic Hamiltonian constraint. Besides
the \emph{motion reversal transformation} which appears naturally
in ordinary (non constrained) classical and quantum systems, we
have described the so called \emph{clock reversal transformation}.
This symmetry arises as a consequence of the existence of a
Hamiltonian constraint with a two sheet structure, being this
symmetry associated to the passage from one sheet to the other
one. We have conjectured that this symmetry could play an
important role for the clarification of the boundary conditions to
be imposed on the space of solutions of the \wdw\ equation.

\section{Appendix A}

The condition for a function $t(q^i,p_i)$ to be a global phase
time (that its Poisson bracket with the Hamiltonian constraint is
positive definite) can be understood as follows. If we define the
Hamiltonian vector field
\begin{equation}
{\rm H}^A =  ({\rm H}^q,{\rm H}^p)
 =  \left({\partial {\cal H}\over \partial p},-{\partial {\cal H}\over\partial
 q}\right),
\end{equation}
then the condition
$$[t,{\cal H}]>  0$$
is equivalent to
$${\rm H}^A{\partial t\over \partial x^A}>  0, $$
with $x^A=(q^i,p_i)$. This means that $t(q^i,p_i)$ monotonically
increases along any dynamical trajectory. Each surface $t= \,
constant$ in the phase space is then crossed only once by any
dynamical trajectory (so that the field lines of the Hamiltonian
vector field are open). The successive states of the system can
then be parametrized by $t(q^i,p_i)$.

If we explicitly write the constraint, the condition for the
existence of a global time which depends only on the coordinates
({\it intrinsic time}) reads
\begin{equation}[t(q^i),{\cal H}]  =  [t(q^i), G^{ik}p_ip_k]
=  2{\partial t\over\partial q^i} G^{ik}p_k>0.\end{equation}

If the supermetric has a diagonal form and one of the momenta
vanishes at a given point of phase space, then no function
depending only on its conjugated coordinate can be a global phase
time. For a constraint whose potential can be zero for finite
values of the coordinates, the momenta $p_k$ can be all equal to
zero at a given point, and $[t(q^i),{\cal H}]$ can vanish. Hence
an {\it intrinsic time} $t(q^i)$ can be identified only if the
potential in the constraint has a definite sign. In the most
general case a global phase time should be a function which
depends also on the canonical momenta; in this case it is said
that the system has an {\it extrinsic time} $t(q^i,p_i)$, because
the momenta are related to the extrinsic curvature (see, for
example \cite{ba93}).

It must be noted that scaling a given constraint by a positive
function does not affect the definition of time, that is, if
$[t,{\cal H}]>0$, then for $H=F(q)\cal H$, $F(q)>0$ we also have
$[t,H]>0$ (we have used this property many times in these notes).
From the point of view of quantization, however, things are less
trivial, and the question arises about the validity of scaling the
Hamiltonian. In this sense, it is important to recall that it can
be shown that  an operator ordering exists such that both
constraints $H$ and ${\cal H}$ are equivalent at the quantum
level. Let us consider a generic constraint
$$
{\cal H}=e^{bq_1}\left(-p_1^2+p_2^2+\zeta e^{aq_1+cq_2}\right)= 0,
$$
which contains an ambiguity associated to the fact that the most
general form of the first term should be written
$$-e^{A q_1} p_1e^{(b-A-C) q_1} p_1e^{C q_1}$$ (so that $A$ and $C$ parametrize all possible operator orderings).
It is simple to verify that the constraint with the most general ordering differs
from that with the trivial ordering in two terms, one linear and one quadratic in
$\hbar$, and that these terms vanish with the choice $C=b-A=0$. Therefore,
the Wheeler--DeWitt equation resulting from the scaled constraint
$H=e^{-bq_1}{\cal H}=0$ is right in the sense that it corresponds
to a possible ordering of the original constraint ${\cal H}$.

\section{Appendix B}

Start from the parametrized action
\be
 S[q^i,p_i,N]=\int_{\tau_1}^{\tau_2}\left( p_i{dq^i\over d\tau }-N{\cal H}\right)
 d\tau\ee
where ${\cal H}=0$ on the constraint surface. Consider the
$\tau-$independent Hamilton--Jacobi equation \be
 G^{ij}{\partial W\over\partial q^i} {\partial W\over\partial q^j}+V(q)=E\ee
which results by substituting $p_i=\partial W/\partial q^i$ in the
Hamiltonian. A complete solution $W(q^i,\alpha_\mu ,E)$ (see for
example \cite{dau60}) obtained by matching the integration
constants
 $(\alpha_\mu ,E)$ to the new momenta $(\overline P_\mu ,\overline P_0)$ generates a canonical transformation
\be
p_i={\partial W\over\partial q^i},\ \ \ \ \ \ \overline Q^i={\partial W\over\partial \overline P_i},\ \ \ \ \ \  \overline K=N\overline P_0=N{\cal H}\ee
which identifies the constraint ${\cal H}$ with the new momentum $\overline P_0$. The variables $(\overline Q^\mu,\overline P_\mu)$  are conserved observables because $[\overline Q^\mu,{\cal H}]=[\overline P_\mu,{\cal H}]=0$, so that they are not    suitable  for  characterizing the dynamical evolution. A second transformation generated by the function
\be
F=P_0\overline Q^0+f(\overline Q^\mu ,P_\mu ,\tau) \ee
gives
\be\overline P_0=P_0\ \ \ \ \ \ \  \overline P_\mu={\partial f\over \partial\overline Q^\mu}\ \ \ \ \ \ \ \
 Q^0=\overline Q^0\ \ \ \ \ \ \  Q^\mu={\partial f\over\partial P_\mu}\ee
and a new non vanishing Hamiltonian
$
K=NP_0+\partial f/\partial\tau,$
so that $(Q^\mu,P_\mu)$ are non conserved observables because $[Q^\mu,{\cal H}]=[P_\mu,{\cal H}]=0$ but $[Q^\mu, K]\neq 0$ and $[P_\mu,K]\neq 0$;  we have,  instead, that $[Q^0,{\cal H}]=[Q^0,P_0]=1$, and then $Q^0$ can be used to fix the gauge.
The  transformation $(q^i,p_i)\  \to\ (Q^i,P_i)$ leads to the action
\be
{\cal S}[Q^i,P_i,N]=\int_{\tau_1}^{\tau_2} \left( P_i{dQ^i\over d\tau}-NP_0-{\partial f\over\partial\tau}\right) d\tau\ee
which  contains a linear and homogeneous constraint $P_0= 0$ and a non zero (true) Hamiltonian $\partial f/\partial\tau$.  Then we have obtained the action  of an ordinary gauge system. Fixing the gauge on this system defines a particular foliation of space-time for the associated cosmological model originally  described by the parametrized action.

In terms of the original  variables the action  $\cal S$ reads
\be
{\cal S}[q^i,p_i,N]= S[q^i,p_i,N]+\left[\overline Q^i\overline P_i -W+Q^\mu P_\mu-f\right]^{\tau_2}_{\tau_1}\ee

We see that  $\cal S$ and $S$ differ only in end point terms; thus
both actions yield the same dynamics. (For a review about the application to the deparametrization and also the subsequent path integral qunatization of minisuperspaces, see Ref. \cite{taty2}).

\end{document}